# Transferable Deep Learning Potential Reveals Intermediate-Range Ordering Effects in LiF-NaF-ZrF$_4$ Molten Salt

Rajni Chahal[1*], Santanu Roy[2], Martin Brehm[3], Shubhojit Banerjee[1], Vyacheslav Bryantsev[2], Stephen T. Lam[1]

[1]Chemical engineering, University of Massachusetts Lowell, Lowell, MA-01854, United States

[2]Chemical Science Division, Oak Ridge National Laboratory, Oak Ridge, TN-37830, United States

[3]Martin-Luther-Universität Halle-Wittenberg, Halle (Saale), Germany

*Supporting Information Placeholder*

**ABSTRACT:** LiF– NaF– ZrF$_4$ multicomponent molten salts are promising candidate coolants for advanced clean energy systems owing to their desirable thermophysical and transport properties. However, the complex structures enabling these properties, and their dependence on composition, is scarcely quantified due to limitations in simulating and interpreting experimental spectra of highly disordered, intermediate-ranged structures. Specifically, size-limited ab-initio simulation and accuracy-limited classical models used in the past, are unable to capture a wide range of fluctuating motifs found in the extended heterogeneous structures of liquid salt. This greatly inhibits our ability to design tailored compositions and materials. Here, accurate, efficient, and transferable machine learning potentials are used to predict structures far beyond the first coordination shell in LiF– NaF– ZrF$_4$. Neural networks trained at only eutectic compositions with 29% and 37% ZrF$_4$ are shown to accurately simulate a wide range of compositions (11 to 40% ZrF$_4$) with dramatically different coordination chemistries, while showing a remarkable agreement with theoretical and experimental Raman spectra. The theoretical Raman calculations further uncovered the previously unseen shift and flattening of bending band at ~250 cm$^{-1}$ which validated the simulated extended-range structures as observed in compositions with higher than 29% ZrF$_4$ content. In such cases, machine learning-based simulations capable of accessing larger time- and length-scales (beyond 17 Å) were critical for accurately predicting both structure and ionic diffusivities.

Molten salts have promising applications in advanced clean energy systems such as next-generation nuclear reactors, solar-thermal storage plants, advanced batteries, and media for carbon capture due to their desirable heat transfer properties. Fluoride salts have been previously identified as good candidates for primary coolant applications in the advanced high temperature reactor (AHTR) [1] and molten salt reactors (MSR) [2][3]. Among them, ternary systems containing BeF$_2$ and ZrF$_4$ were recommended as coolant salts. While there has been a significant interest in using BeF$_2$-based salts due to their low neutron absorption, there remain substantial challenges in using beryllium salts due to their toxicity and required processing facilities. As such, Zr-salts present a compelling alternative due to their acceptable neutron economy, vapor pressures, thermal hydraulics, and lower costs [3]. In order to achieve low vapor pressure at higher temperatures (<1 mm Hg at 700-900°C), the ZrF$_4$ mole fraction in the salt mixture should be maintained within ~20-45% [3]. Here, the eutectic compositions 26LiF – 37NaF – 37ZrF$_4$ (mol%) and 42LiF – 29NaF – 29ZrF$_4$ (mol%) with freezing point around 436°C and 460°C, respectively have been identified as promising candidates. Due to lack of data in thermophysical properties databases [4], these salt compositions have been recommended for further study [2].

Precise experimental interrogation of salt structure and properties is challenged by the radioactive environment, high temperature conditions, cost, materials handling, and difficulties in interpreting experimental data. Particularly, it is difficult to access the structure of multivalent cations using techniques such as Raman spectra and extended X-ray absorption fine structure (EXAFS) alone due to their existence in multiple coordination states and intermediate-range ordering [5][6][7]. Specifically, deducing precise local structures from ensemble-averaged Raman spectra is challenging when there is high dynamic disorder and heterogeneity in coordination environments [8]. Likewise, concerning the intermediate-range structure, a previous experimental Raman spectroscopic study by Toth *et al.* suggested no fluorozirconate chain formation, a.k.a., intermediate-range ordering, even for the composition involving 40% mole content of ZrF$_4$ [9]. However, while the local coordination states reported in this Raman study was found to be in good agreement with our previous ab-initio molecular dynamics (AIMD) findings, AIMD simulations for salts with 29% and 37% ZrF$_4$ content revealed fluorozirconate chain formation [10]. Such intermediate-range structural ordering effect was also previously reported from nuclear magnetic resonance (NMR) and EXAFS experimental observations and classical molecular modeling for LiF– ZrF$_4$, NaF– ZrF$_4$, and KF– ZrF$_4$ salts where higher ZrF$_4$ mole content was present [11][12].

In light of the challenges encountered in experimental measurements and their interpretation, AIMD simulations can be used to interpret experimental data, and predict temperature-dependent structure, transport, and thermophysical properties of multicomponent molten salts [13][14]. However, accurately simulating the structure and properties of multicomponent molten salt containing multivalent cations still remains challenging due to fundamental limitations in electronic structure methods based on density functional theory (size-limited). As such, the classical molecular dynamics (CMD) simulations have been proven efficient when studying the larger system sizes. However, our previous study showed that the traditional Rigid ion model (RIM) parameters overpredict the zirconium coordination numbers as well as the intermediate-range ordering when ZrF$_4$ mol% is higher in the melt, which in turn lead to up to 2-3 orders of magnitude differences in the diffusion coefficients and viscosities values [10]. Such issues at higher ZrF$_4$ mol% content can be essentially fixed by including the polarization effects leading to the development of Polarizable ion model (PIM) [15]. Here, this addition of polarization term requires accurately capturing the charge-dipole and dipole-dipole polarizability [16], which has seen many applications in the simulation of multicomponent salts [15][17][18][19][20].

While the fitted PIM parameters were used by Salanne et al. [21] to study electrical conductivities of molten LiF–NaF–ZrF$_4$ mixtures, a detailed study to validate the PIM-generated salt structure and transport properties using AIMD and experimental data is yet to be reported. Even though the transferability of classical interatomic potentials is an attractive trait, their further development for material screening is often overshadowed by the tedious challenges in parameter fitting, namely, ensuring excellent quality fit for both force and dipole values, the quality of experimental and first-principles data used for parameter optimization, and excessive human intervention [15]. Here, density functional theory (DFT)-accurate as well as nearly CMD-efficient neural network interatomic potentials (NNIPs) can overcome these limitations and can be trained directly on the AIMD data without requiring any significant human intervention in defining and fitting parameters that are extensible to arbitrarily complex systems. Previously, Lam et al. have shown the robustness and versatility of NNIP when employed to study multicomponent molten salts containing multivalent cation species (Be$^{2+}$) [22]. Along this line, Rodriguez et al. [23] used DeePMD-kit [24][25] to develop an NNIP to generate an accurate salt structure for the prediction of transport and thermophysical salt properties. As the properties such as diffusion coefficients, conductivity, and viscosity of the salt melt, etc. are strongly influenced by the formation of coordination complexes, their lifetime, and the degree of their connectivity (chain formation) [20], the NNIP transferability across different salt compositions entail the requirement for accurate prediction of short to intermediate-range structure.

It is previously shown that greater is the complexity of the machine learned potentials, such as in Neural network (NN) and Gaussian approximation potential (GAP), the greater is the issue with their transferability when deployed outside training thermodynamic conditions [26]. To enhance their transferability, previous studies suggested implementing an active learning loop based on the desired uncertainty quantification approach [27][28][29][30]. This further increases the complexity of NNIP development in addition to increasing the count of expensive AIMD calculations in the regions of higher uncertainty. In this letter, we present a systematic approach for the development of NNIP trained in the limited phase-space of LiF-NaF-ZrF$_4$ salt that can accurately reproduce the short-range coordinated complexes and intermediate-range ordering effects across a wider phase-space. The fitted NNIP is used to study the short to intermediate range salt structure of five compositions of LiF–NaF–ZrF$_4$ salts [31]: 38–51–11 (A), 40–46–14 (B), 42–29–29 (C), 26–37–37 (D) and 28–32–40 (E) mole% of LiF–NaF–ZrF$_4$ at 750°C, 650°C, 727°C, 700°C, and 550°C, respectively. The NNIP training was performed using VASP-generated data for composition C and D at 1000K and 973 K, respectively using DeePMD-kit [24]. The NNIP training data was selected following the recommendations made by Rodriguez et al. for Flibe NNIP development [22]. A detailed description of the training dataset and network parameters can be found in the SI. The trained NNIP potential was then used to study the structure and ionic diffusivities for compositions A (11% ZrF$_4$), B (14% ZrF$_4$), and E (40% ZrF$_4$) in which the ZrF$_4$ content, and consequently degree of structural ordering, is outside the range of the training dataset (29% and 37% ZrF$_4$).

Specifically, the short to intermediate-range structure of LiF-NaF-ZrF$_4$ salt is explored using AIMD (plane wave, DZVP, TZV2P), PIM, and NN-based molecular dynamics (NNMD) simulations. The effect of increasing ZrF$_4$ content on the salt structure is studied.

Thereafter, the appropriate cell size for each composition is explored to accurately capture the intermediate-range ordering effects in compositions containing higher ZrF$_4$ mol%. Further, the effect of cell parameter on the structural ordering and transport properties (self-diffusion coefficients) of the salt is discussed. The coordination states and structural ordering is validated using experimental and theoretical Raman spectral analysis. Here, an updated interpretation of experimental Raman spectra is provided based on the observed fluorozirconate chain formation for higher ZrF$_4$ content compositions in our simulations.

The trained NNIP potential is first used to study the short to intermediate range salt structure, which is validated by the Raman spectral observations. The cation-anion coordination numbers are evaluated and compared within the NNMD, PBE-D3 VASP (plane wave), Quickstep/CP2K (DZVP, TZV2P) and PIM simulations. Additional details of LAMMPS MD simulations [32][33], DFT calculations [23][34][35], and CP2K calculations [36][37] are provided in SI. All methods agree well with the observations made in the Raman study which confirmed the occurrence of predominately 6, and 7 coordinated fluorozirconate complexes with a relatively small amount of 8-coordinated complexes (**Fig 1b, Fig. S2**). However, the observed trend for average fluoride coordination number of Zr (CN) in PIM simulations deviates from as suggested by shift in Raman spectra, whereas both plane wave and NNMD simulations closely capture this trend (**Fig. 1a**). In general, the shift to higher frequency in Raman spectra indicates a decrease in the average F-Zr CN upon increasing the ZrF$_4$ content. This trend is generally predicted by the AIMD (plane wave, TZV2P, DZVP) and NNMD simulations for all compositions except B. The observed exception in CN trend can be explained from the perspective of difference in simulation temperatures. It is known that temperature plays a pivotal role in the F-Zr coordination number (CN increases as temperature decreases) [38]. Here, the comparatively lower temperature for composition B is responsible for slight increase in average Zr CN for composition B, which only has a slightly higher ZrF$_4$ content than composition A. Going from composition D to E, plane wave and NNIP simulations show a decrease in average Zr CN (in agreement with Raman spectra observation), while CMD predicted value suggests a slight increase in average Zr CN. This anomaly in average Zr CN prediction using PIM parameters was also previously reported by Pauvert et al. for LiF-ZrF$_4$ system [38].

In addition to the NNIP's ability to accurately predict the average CN, the NNMD predicted population of the 6-, 7-, and 8-fold-fluorozirconate complexes show better agreement with AIMD (plane wave, TZV2P, DZVP) when compared to CMD computed values. The stability of the different coordination states in the melt can be further explored using free energy calculations. Here, while **Fig. 1** reveals the average coordination environment, **Fig. 2** illustrates distributions and metastability of different coordination structures in terms of their free energies (see SI, Equation S2, for definition). All simulations indicate that the fluoride coordination number of a zirconium ion is dominated by 6 and 7 while 8 is less likely to occur. The AIMD data suggests that the composition with 14 mol% Zr (Composition B) prefers 7 coordinate state more and requires to overcome a high barrier of 6 kcal/mol to transit to 6 coordinate state. On the other hand, the composition with 40 mol% Zr (Composition E) prefers 6-coordinate state slightly more than the 7-coordinate state and the barriers between them are much smaller (2-2.5 kcal/mol) indicating their frequent interconversions. The NNIP data agree well with this preference of coordination number of Zr, even though the change in the Zr mol% is rather less drastic from Composition C (29 mol%) to D (37 mol%) to E (40 mol%). The PIM data do not exhibit this sensitivity. For zirconium coordination number of a fluoride ion all simulations agree with the distributions of coordination numbers and their relative stabilities—the 1-coordinate state is the most-likely state while the 2-

coordinate state is like to occur with increasing mol% of Zr. Here, the observed existence of 2-coordinate state corresponds to the sharing of fluorine ion among two fluorozirconate complexes leading to the intermediate-range ordering effects.

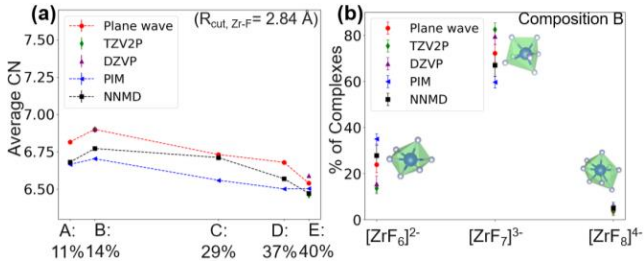

**Fig. 1** NNIP Transferability and comparison of local structure via (a)Average F-Zr coordination number, (b)population of fluorozirconate complexes in composition B.

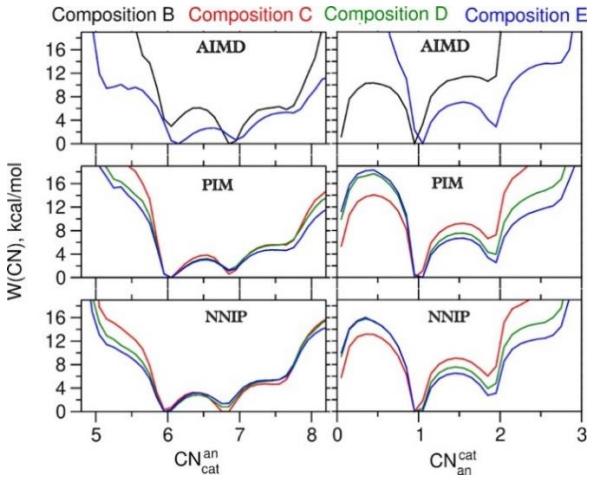

**Fig. 2** Free energy profiles highlighting the distributions of fluoride coordination number of $Zr^{4+}$ (left) and zirconium coordination number of $F^-$ for different composition with creasing zirconium content (black to blue) obtained from different molecular dynamics simulations.

In order to explore such intermediate-range structure formation in LiF-NaF-ZrF$_4$ in training (**Fig. 3**) as well as test (**Fig. 4**) regime, Zr-Zr RDF, Zr-F-Zr angular distribution, and structure visualization (**Fig. 3, 4** inset) are employed. When analyzing the Zr-Zr RDF, the first peak at ~4 Å represents the $([ZrF_x]^{4-x})_n$ chain formation due to edge- and corner-sharing $[ZrF_x]^{4-x}$ complexes. As no first peak is observed for compositions A and B, the salt structure mainly comprises of isolated 6-, 7-, and 8-coordinated $[ZrF_x]^{4-x}$ complexes (**Fig. 4a, b** inset). As the ZrF$_4$ mole% increases, the first peak in Zr-Zr RDF becomes more significant representing an increase in fluorozirconate chain formation (**Fig. 3a, b** inset). All the simulations methods used in this study agree well on predicting the Zr-Zr RDF at lower ZrF$_4$ content. However, in the cases where $([ZrF_x]^{4-x})_n$ chain formation is observed (compositions C, D, E), the intermediate-range ordering is more accurately predicted by NNMD when compared to the PIM simulation. For composition C, PIM underpredicts the $([ZrF_x]^{4-x})_n$ chain formation (**Fig. 3a**), which is also evident from the lower count of Zr-F-Zr in ADF (**Fig. 3c**). As ZrF$_4$ mole% further increases (composition C & D), PIM generated structure is more dominated by the edge-sharing complexes, as observed from the Zr-F-Zr angles and the shift to the left in the first peak in Zr-Zr RDF. The dominant edge-sharing chain connection can be attributed to the higher F$^-$ polarizability effect captured by PIM, which allows more shielding of coulombic repulsive interactions between Zr-Zr from the adjacent complexes in the chain. This, in turn, decreases the Zr-F-Zr angle which leads to the decrease in Zr-Zr distance (left shift in first peak in Zr-Zr RDF). We have previously found that omitting the polarization contribution results in the chain primarily connected via corner-sharing complexes following the same reasoning provided above [10]. Here, NNMD is able to accurately predict the fluorozirconate chain connected via both corner-sharing and edge-sharing complexes for both training and test compositions.

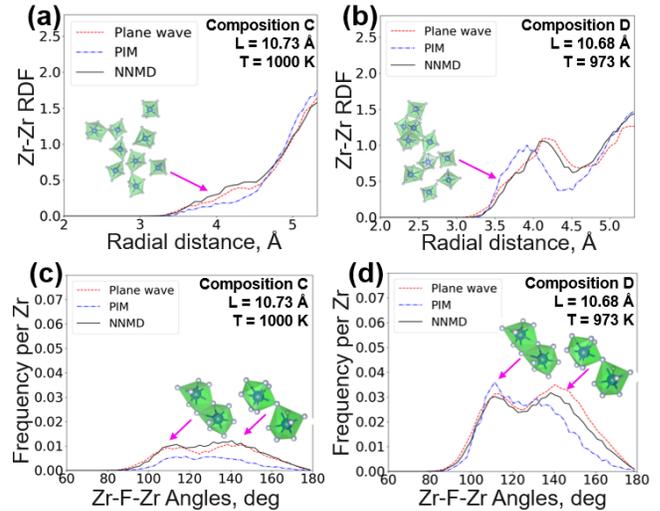

**Fig. 3** Comparison of NNMD-predicted (a, b) RDF and (c, d) ADF with AIMD and PIM simulations for compositions in training regime. The snapshots of the corresponding representative salt structure are shown next to the Zr-Zr first peak in Zr-Zr RDF.

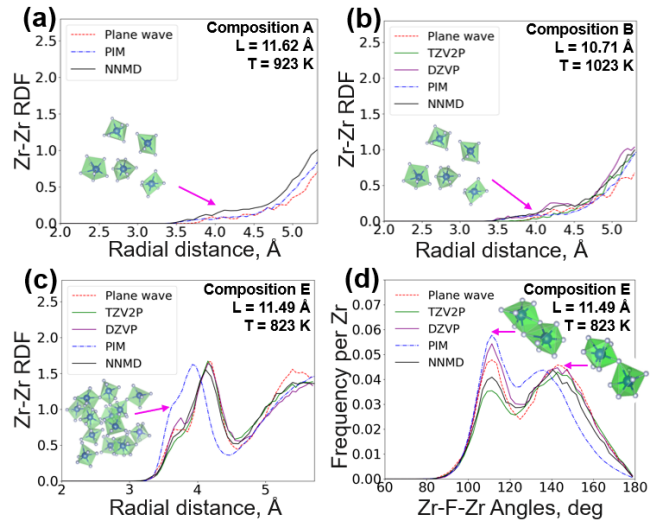

**Fig. 4** NNIP Transferability and comparison of intermediate-range structure ordering across a wide phase-space via (a, b) Zr-Zr RDF, and (c, d) Zr-F-Zr ADF. The snapshots of the corresponding representative salt structure are shown next to the Zr-Zr first peak in Zr-Zr RDF.

The predicted salt structure in the simulations can be directly validated via Raman spectroscopy [39][40][41], which has been traditionally employed to probe the local structure of molten salts and to understand how the structure of melts changes by varying composition and temperature. Spectral shifts in the Raman bands by varying ZrF$_4$ from 14 to 40% in the early work of Toth et al. were interpreted as changes in the preferred coordination of zirconium from eight to five with no fluoride bridging between Zr polyhedra. This picture was challenged and substantially refined in a more recent work of Papatheodorou and co-workers [7]. From the Raman spectra of ZrF$_4$-KF mixtures and related compounds they inferred a two species equilibrium between the seven- and six-coordinated zirconium ions

$$ZrF_7^{3-} \rightleftharpoons ZrF_6^{2-} + F^-,$$

and a propensity of forming more extended chain structures in melts rich in ZnCl$_4$. The AIMD and NNMD generated structures in molten salts containing mostly seven- and six-coordinate zirconium complexes (**Fig. 1** and **2**) support the proposed equilibrium. The simulated Raman spectra (**Fig. 5**) generated from the TZV2P basis set agree well with the experimental spectra, both in terms of the overall shape and the shift of the main totally symmetric stretching bands to higher frequencies with increasing ZrF$_4$ concentration. The latter is a consequence of decreasing coordination number and increasing fluoride bridging, in which the Zr-F$_t$ frequency with the terminal fluoride anions (F$_t$) is blue shifted relative to the band of the monomeric species [7][42]. Besides the shift of the main peaks, simulations reproduce the flattening of the bending band at ~250 cm$^{-1}$ and the shift to lower frequencies in going from 14 to 40% ZrF$_4$. This is a direct result of inhomogeneous broadening associated with the depletion of the intensity coming from pure monomer species and a superposition of multiple bands at lower frequencies due to multiple chains formed by zirconium polyhedra. A shoulder in the experimental spectrum or a broader spectral feature in the simulated spectrum at ~330 cm$^{-1}$ in dilute ZrCl$_4$ was assigned to the E$_2'$ vibration mode of the pentagonal bipyramidal ZrF$_7^{3-}$ species [7]. Underestimation of the absolute positions of the experimental bands, which is < 30 cm$^{-1}$ in all cases, is presumably due to a slight overprediction of the stability of ZrF$_7^{3-}$ over ZrF$_6^{2-}$. This underscores high sensitivity of the position of the Raman bands to changes in the coordination, providing useful experimental metrics to benchmark the accuracy of DFT in describing local structure of molten salts.

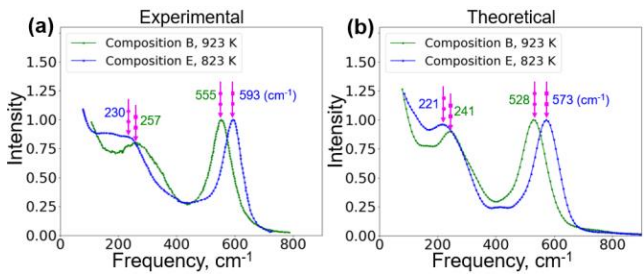

**Fig. 5** (a) Experimental vs. (b) simulated unpolarized Raman spectra for compositions B and E generated based on the AIMD trajectories using the TZV2P basis set. All Raman spectra are normalized for the main totally symmetric stretching band.

Given the formation of extended range structures as observed from simulations and interpretation of Raman spectra, the periodic boundary conditions in small simulation cell sizes can result in structural artifacts. Here, the trained NNIP potential is used to investigate the influence of simulation cell parameter on the structure and transport properties of each composition. The simulation cell size influence on the local structure is investigated using the average value of Zr CN as well as the population of existing coordination states. We found that as the cell size reaches ~ 12.43 Å (C), the average Zr CN and population of large [ZrF$_x$]$^{4-x}$ complexes increases and remained unchanged upon further increasing the cell size (**Fig. 6, Fig. S3**). This trend follows as observed from the plane wave AIMD values for composition C and D (**Fig. 6a, Fig. S3**). It can be attributed to the improved averaging over an increased number of Zr complexes as the cell size increases. The inaccurate population of different coordination complexes due to restrictions on cell size and simulation trajectory length has been previously discussed in other AIMD studies [43].

In order to explore the effect of cell parameter on the intermediate-range structure of LiF-NaF-ZrF$_4$, the Zr-Zr RDF, Zr-F-Zr angles distribution, and fluorozirconate cluster size ((([ZrF$_x$]$^{4-x}$)$_n$) distribution are employed. When lower ZrF$_4$ mole % is present (compositions A and B), the Zr-Zr RDF essentially remains unchanged as the cell size increases (**Fig. 7a & b**). Here, as no ([ZrF$_x$]$^{4-x}$)$_n$ cluster formation occurs, the corresponding Zr-F-Zr ADF plots are not shown. Further, as the ZrF$_4$ mole% increases, the first peak in the Zr-Zr RDF rapidly rises with the increase in cell size ~10.73-12.43 Å (C) and 10.65-13.45 Å (D). This indicates an increase in the formation of more ([ZrF$_x$]$^{4-x}$)$_n$ clusters, which were previously restricted due to a smaller simulation cell lengths. This increase in chain formation upon larger cell length is also reflected from the increase in the frequency in Zr-F-Zr angles distribution, where the area under the curve is proportionate to the ([ZrF$_x$]$^{4-x}$)$_n$ clusters formed. Here, even though the ([ZrF$_x$]$^{4-x}$)$_n$ cluster formation changes upon change in cell parameter, the chains stay connected via both edge- and corner-sharing complexes. The convergence of Zr-Zr RDF and Zr-F-Zr angles distribution past cell sizes ~12.43 Å (C), 13.45 Å (D), and 14.47Å (E) implies these cell sizes to be the sufficient for the respective compositions (**Fig. 7**). As both the Zr-Zr RDF and Zr-F-Zr angles unfold the structure up to second coordination shell while providing only the system average, a further investigation is required to identify and quantify the ([ZrF$_x$]$^{4-x}$)$_n$ clusters sizes and their distribution.

In addition to dimers, we identified the ([ZrF$_x$]$^{4-x}$)$_n$ clusters of various sizes throughout the equilibrated trajectory by searching for their connectivity based on the cut-off value equal to Zr-F bond length (minima in Zr-F RDF = 2.84 Å). The probability of finding ([ZrF$_x$]$^{4-x}$)$_n$ clusters of sizes greater than 2 for compositions C, D, and E is plotted in **Fig. 8**. Here, we noticed that even though both Zr-Zr RDF and Zr-F-Zr angles distribution converges for cell sizes ~12.43 Å (C), 13.45 Å (D), and 14.47Å (E), the distribution of ([ZrF$_x$]$^{4-x}$)$_n$ clusters keep developing until cell size reaches more than nearly 17 Å (C), 21 Å (D), and 23 Å (E). Therefore, while the comparatively smaller simulation cell size is sufficient to capture the accurate salt structure in the cases where ZrF$_4$ is low, the cell size hinders the representation of intermediate-range structural ordering when more ZrF$_4$ content is present (as is the case with reactor-relevant compositions)**.**

As the intermediate-range ordering has been previously reported to affect the transport properties like diffusion coefficients and viscosities [20][10], it is crucial to choose the appropriate cell parameter based on the convergence in intermediate structure ordering. It is also shown in our study by evaluating the diffusion coefficients for each cell size for the considered salt compositions using block-diffusivity method [44]. Notably, as the smaller cell sizes were sufficient to accurately represent the relatively simplified salt structure (isolated fluorozirconate complexes: [ZrF$_x$]$^{4-x}$) in composition A and B, the calculated diffusion coefficients are insensitive to the increase in the cell parameter (**Fig. 9a & c, Fig. S4, Fig. S5**). On the contrary, when a significant structural ordering effects are present, the ionic diffusion coefficients are found to vary by up to 130% as the cell sizes increase (**Fig. 9b & d, Fig. S6, Fig. S7, Fig. S8**). As soon as a representative cell size is attained for each composition, a plateau is observed in the values of diffusion coefficients, which agrees with our observations from ([ZrF$_x$]$^{4-x}$)$_n$ cluster distribution. The observed strong agreement in the trend in convergence of different cluster sizes and the diffusivity values upon reaching a certain cell size emphasizes the significance of accurately predicting intermediate-range fluorozirconate structures towards accurate prediction of ionic diffusivities. Therefore, deciding on the appropriate cell size solely based on second coordination shell analysis (Zr-Zr RDF and Zr-F-Zr angles) would result in inaccurate self-diffusivity values. Here, it should be emphasized that such an exploration of appropriate cell sizes, and calculation of long trajectories for equilibrium properties would not be feasible solely from the AIMD simulations due to limitations on practical cell size and simulation times. It can be further noticed that as the $ZrF_4$ mole% vary from 11% (A) to 29% (C), the self-diffusivity values drastically decrease. This is due to the formation of previously discussed intermediate-range structure motifs $(([ZrF_x]^{4-x})_n)$ as $ZrF_4$ mole% increases. Further, the self-diffusivity values consistently decrease as the $ZrF_4$ mole% vary from 29% (C) to 40% (E), which can be attributed to the increase in the count of large fluorozirconate chains (e.g., n >=7) as $ZrF_4$ content increase from 29% to 40% (**Fig. 9**). Such increase in intermediate-range structural ordering is also evident from the increase in the height of first peak in Zr-Zr RDF (**Fig. 7c, e, g**) as $ZrF_4$ content increases. Overall, the observed trend in diffusivities and fluorozirconate chain formation indicates a direct relationship between the intermediate-range structural ordering and transport properties. Such differences in structure and diffusivity may further contribute to the differences in other important transport and thermophysical properties. In this respect, the effect of limited cell size on the density, heat capacity, thermal conductivity, and viscosity needs to be further studied using the developed NNIP potential.

**Fig. 6** Exploring cell size effect of short-range structure using NNMD via (a) Average Zr CN (b) population of $[ZrF_x]^{4-x}$ complexes in composition C.

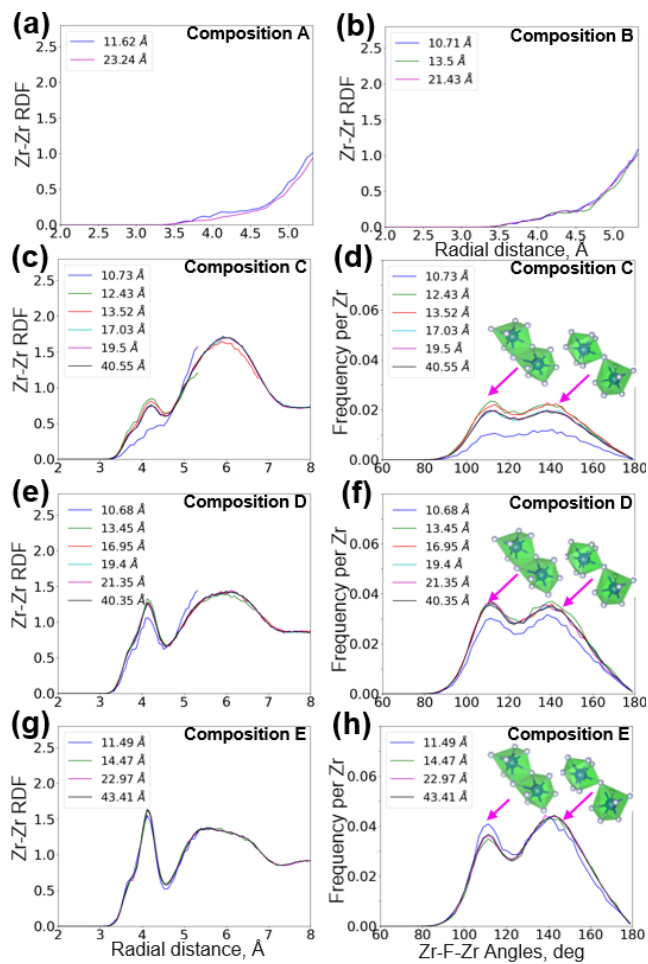

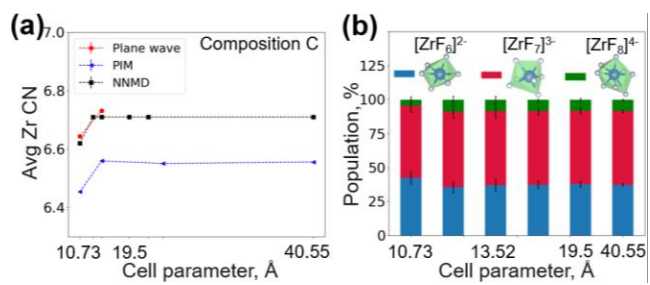

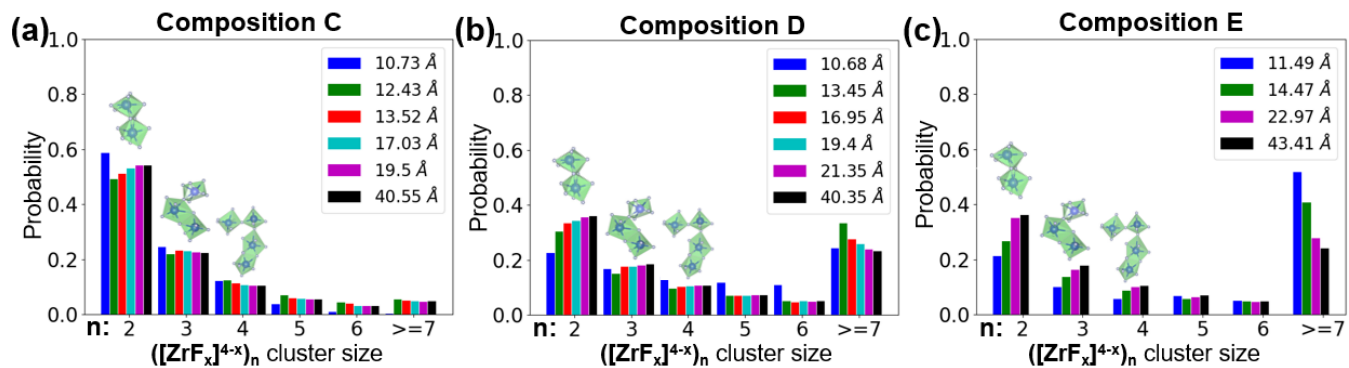

**Fig. 7** Exploring cell size effect on intermediate-range ordering using NNMD via (a,b,c,e,g) Zr-Zr RDF and (d,f,h) Zr-F-Zr angles distribution. Fig. S9 shows full RDFs from large box simulations.

**Fig. 8** Appropriate cell size exploration using $([ZrF_x]^{4-x})_n$ cluster size distribution using NNMD simulations. The snapshots of cluster size 2, 3, and 4 are shown to guide the reader. Fig. S10 shows snapshots of all cluster sizes. The $x$ corresponds to 6, 7, and 8 coordination state of individual fluorozirconate complex ($[ZrF_6]^{2-}$, $[ZrF_7]^{3-}$, $[ZrF_8]^{4-}$).

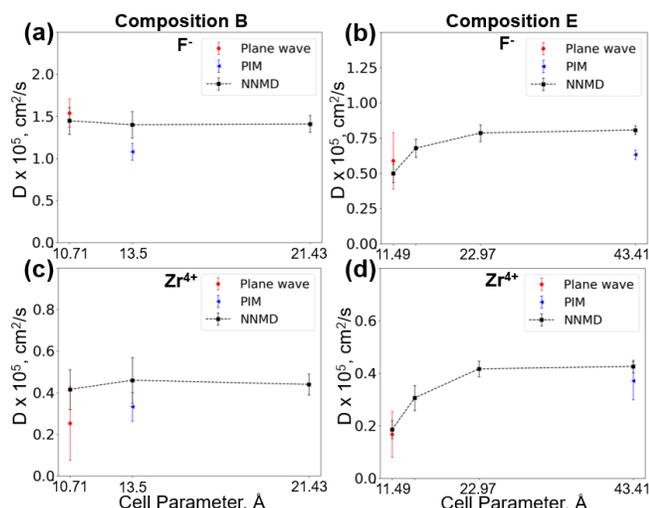

**Fig. 9** Effect of cell parameter on self-diffusivity using NNMD when (a) isolated $[ZrF_x]^{4-x}$ exist (composition B) and (b) extended-range structures are formed (composition E).

Overall, this work explored the existence of structural ordering in multicomponent molten salts that is often overlooked in understanding transport property trends. Specifically, the intermediate-range ordering effects in the LiF-NaF-ZrF$_4$ salt were explored using the NNIP potential trained on 29% and 37% ZrF$_4$ compositions. It was found that as mole% of multivalent cation species (ZrF$_4$) increases in the melt, the extended fluorozirconate chain clusters were formed, which resulted in a commensurate decrease in the ionic diffusion. Further, when such extended structures exist in the melt, a simulation cell parameter of at least 17 Å was found to be critical to correctly represent the extended-range structure as well as the ionic diffusivities. The increased simulation cell sizes, improved sampling, and DFT-level accuracy of developed NNIP allowed for an accurate prediction of intermediate-range structure and ionic diffusivities, which consistently surpassed the CMD predictions obtained using physics-based polarizable ion models. Besides the extended-range structure prediction, the developed NNIP showed excellent agreement for average coordination and the stability of 6-, 7-, and 8-coordinated fluorozirconate species when compared to ab-initio calculations and Raman spectral observations. In addition to validating the NNMD predicted structure, the theoretical Raman spectral calculations also revealed the shift and flattening of the bending band at ~250 cm$^{-1}$, which correspond to the structural ordering effects previously misinterpreted in the experimental spectral observations. As such, the demonstrated transferability of NNIP across a wide range of compositions (11% to 40% ZrF$_4$) and thermodynamics conditions opens up the possibility of the development of robust and reliable neural network potentials to advance the screening of new and unseen chemical compositions of many structurally-complex liquids.

## ASSOCIATED CONTENT

### Supporting Information

To be updated.

## AUTHOR INFORMATION

### Corresponding Author

rajni_chahal@uml.edu; stephen_lam@uml.edu

### Author Contributions

To be updated.

### Notes

The authors declare no competing financial interests.

## ACKNOWLEDGMENT

This work is supported by DOE-NE's Nuclear Energy University Program (NEUP) under Award DE-NE0009204. M.B. acknowledges financial support by the Deutsche Forschungsgemeinschaft (DFG) through project Br 5494/1-3. A part of the computational resources were provided by Massachusetts green high performance computing cluster (MGHPCC). This research also used resources of the National Energy Research Scientific Computing Center, a DOE Office of Science User Facility supported by the Office of Science of the U.S. Department of Energy under Contract No. DE-AC02-05CH11231, awards ASCR-ERCAP0022362 and BES-ERCAP0022445.